# Highlights

**Mineral and cross-linking in collagen fibrils: The mechanical behavior of bone tissue at the nano-scale**

Julia Kamml,Claire Acevedo,David S. Kammer

- Mineral position and morphology influence the mechanics of the collagen fibril.
- At low mineral contents, AGEs cross-links are governing the mechanical response.
- At high mineral contents, the mechanical response is dominated by the mineral.
- Minerals change the fracture mechanics: a high number of collagen molecules ruptures.
- Collagen fibril mechanics are possibly highly adjusted via mineral content.

# Mineral and cross-linking in collagen fibrils: The mechanical behavior of bone tissue at the nano-scale

Julia Kamml[a], Claire Acevedo[b] and David S. Kammer[a,*]

[a]*Institute for Building Materials, ETH Zurich, Switzerland*
[b]*Department of Mechanical and Aerospace Engineering, University of California San Diego, San Diego, California, USA*

ABSTRACT

The mineralized collagen fibril is the main building block of hard tissues and it directly affects the macroscopic mechanics of biological tissues such as bone. The mechanical behavior of the fibril itself is determined by its structure: the content of collagen molecules, minerals, and cross-links, and the mechanical interactions and properties of these components. Advanced-Glycation-Endproducts (AGEs) cross-linking between tropocollagen molecules within the collagen fibril is one important factor that is believed to have a major influence on the tissue. For instance, it has been shown that brittleness in bone correlates with increased AGEs densities. However, the underlying nano-scale mechanisms within the mineralized collagen fibril remain unknown. Here, we study the effect of mineral and AGEs cross-linking on fibril deformation and fracture behavior by performing destructive tensile tests using coarse-grained molecular dynamics simulations. Our results demonstrate that after exceeding a critical content of mineral, it induces stiffening of the collagen fibril at high strain levels. We show that mineral morphology and location affect collagen fibril mechanics: The mineral content at which this stiffening occurs depends on the mineral's location and morphology. Further, both, increasing AGEs density and mineral content lead to stiffening and increased peak stresses. At low mineral contents, the mechanical response of the fibril is dominated by the AGEs, while at high mineral contents, the mineral itself determines fibril mechanics.

## 1. Introduction

Type 2 Diabetes mellitus (T2DM) is widely recognized as a significant contributor to serious health complications in the human body. In addition to adverse effects on the cardiovascular system and other organs such as kidneys and eyes, researchers have found that individuals with T2DM have an increased risk of bone fractures. (Strain and Paldánius, 2018; Milicevic et al., 2008; Prasad et al., 2012; Thomas et al., 2015; Poiana and Capatina, 2017; Oei et al., 2015). The standard clinical metric for assessing fracture risk is bone mineral density (BMD). Typically, decreased bone mineral density (BMD) correlates with conditions like osteoporosis or osteopenia, indicating a higher risk of fractures. The fracture risk in T2DM patients also remains elevated despite exhibiting a normal to slightly elevated BMD (Vestergaard, 2007; Yamamoto et al., 2009; Abdulameer et al., 2012; Montagnani et al., 2011). The potential explanation for this paradox could lie in the high content of Advanced-Glycation-Endproducts (AGEs) in T2DM individuals. AGEs are built via glycosylation within the collagen fibrils of tissue in the presence of sugars, a process also known as the Maillard reaction (Avery and Bailey, 2006). Due to increased glucose levels in the system, the prevalence of these molecules is increased in diabetic patients (Vlassara and Palace, 2002; Vlassara and Uribarri, 2014; Vlassara and Striker, 2013). This increased content of AGEs has been shown to correlate with an increased fracture risk and brittleness in bone at the macro-scale and with deterioration of the collagen fibril's ability to deform at the nano-scale (Yamamoto and Sugimoto, 2016; Acevedo et al., 2018). However, the precise mechanisms through which AGEs influence bone mechanics have not been fully elucidated, including the extent to which AGEs contribute to the impairment of material behavior.

The hierarchical composite structure confers bone with unique material properties that are highly adapted to its function, such as providing support for movement and protection of vital organs (Rho et al., 1998; Ulrich Meyer, 2006; Weatherholt et al., 2012). Apart from the organic constituents, with the collagen fibril as the basic unit, bone also comprises a mineral phase and water. Collagen type I is the most abundant protein in the extracellular matrix of bone, comprising about 95% of the collagen content, which makes it the most important structural unit (Niyibizi and Eyre, 1994; Viguet-Carrin et al., 2006). Its basic building blocks are tropocollagen (TC) molecules bundled up to form the collagen fibrils, with a diameter between 20 to 500 *nm* and a length of about 100 *μm* (Henriksen and Karsdal, 2024; Hulmes, 2008, 2002). TC molecules consist of three coiled peptide helices with non-helical telopeptide areas at each end of the molecule (Hulmes, 2008). In the longitudinal direction, their structural built-up displays the collagen-specific staggered pattern with five gap and overlap zones per TC length (Wess, 2008). AGEs are usually formed at helical regions of the TC molecules, but very little is known about their exact location, which also depends on the AGE type. We distinguish between non-cross-linking AGEs attached to the TC molecules and cross-linking AGEs linking two neighboring TC molecules within the collagen fibril structure. The cross-linking AGEs are suspected of preventing the deformation behavior of the collagen fibril at the nano-scale level of tissue, but the exact mechanisms remain unknown (Fessel et al., 2014; Li et al., 2013; Acevedo et al., 2018; Rosenberg et al., 2023a,b; Vashishth, 2009; Gupta et al., 2006; Zimmermann et al., 2011). Since the collagen fibril is the main building component of bone, alterations in its deformation behavior resulting from elevated AGE contents are expected to influence the behavior of bone on the tissue level. However, the precise mechanisms remain unclear.

*Corresponding author
✉ dkammer@ethz.ch (D.S. Kammer)
orcid(s): 0000-0003-3782-9368 (D.S. Kammer)

In previous studies, it has been shown that AGE cross-linking significantly changes the deformation and fracture behavior of the non-mineralized collagen fibril (Kamml et al., 2023a,b). In particular, it was observed that the strength and stiffness of the fibril increases with increasing AGE densities and AGEs' loading energy capacity. The deformation behavior of the fibril presents a stiffening at high strain levels when the cumulative loading energy capacity of all AGEs exceeds the loading energy capacity of collagen bonds within the collagen molecules. Further, it was demonstrated that changes in the failure mechanism causing this stiffening eventually resulted in a more brittle failure, *i.e.*, energy is rather absorbed via stretching than dissipated via inter-molecular sliding of the collagen molecules in the presence of high AGE contents. This leads to a more sudden energy release when the stretched bonds break. While these results have revealed the origin of brittle failure in non-mineralized collagen fibrils in the presence of AGEs, they do not provide direct and clear evidence for impaired bone tissue behavior since the collagen fibrils in bone are mineralized.

Mineralization of collagen fibrils occurs in a process in which mineral crystals are deposited onto the bone matrix, facilitating the development and strengthening of the bone (Dey and Dey, 2020). The addition of mineral content confers bone with exceptional elastic properties, mostly stiffness, leading to higher strength (Zimmermann and Ritchie, 2015). It has been shown that an increasing amount of mineral correlates with increasing elastic modulus and yield stress (Wachter et al., 2002). The mineral crystals are apatite, similar to hydroxyapatite (HAP) but have a less perfect structure. Still, it is generally referred to as HAP. Different theories exist regarding the exact location and distribution of minerals within bone tissue. It is generally believed that the nucleation of minerals starts in the gap zones of the collagen fibril (Arsenault, 1989; Weiner and Traub, 1992; Tong et al., 2003). Nevertheless, the precise arrangement remains a subject of ongoing debate given that the mineral represents about 65% of hydrated bone weight. However, the gap zones of the collagen fibril do not provide sufficient space to accommodate this large amount of material. It is extremely challenging to extract mineralized collagen fibrils from bone for mechanical tensile testing and the question of how minerals and their distribution and structure influence fibril mechanics on the small scale and tissue on the larger scale has not been answered yet. Some techniques using small-angle X-ray scattering (SAXS) during in situ tensile tests of bone samples have been applied to reveal the loss of collagen deformation capacity in the presence of high AGEs contents (Gupta et al., 2006; Acevedo et al., 2018). In addition to experiments, *in-silico* testing provides a valuable tool for investigating the mechanical behavior of collagen fibrils: Depalle et al. (2015, 2016) and Tavakol and Vaughan (2023) have used coarse-grained molecular dynamics to investigate the influence of different parameters such as enzymatic cross-linking and mineralization on the collagen fibril deformation behavior, but the effect of AGEs in the mineralized fibril has not been investigated so far.

Here, we aim to provide numerical evidence for the influence of both, AGEs cross-linking and minerals, on the mechanics of collagen fibrils. We perform *in-silico* destructive tensile tests on mineralized collagen fibril models with different AGEs densities and mineral contents using coarse-grained steered molecular dynamics simulations. We investigate the influence of mineral content and morphology and AGEs cross-linking on collagen fibril deformation and fracture behavior, revealing the effect of these parameters on the nano-scale mechanics of bone.

## 2. Material and methods

We use a 3D coarse-grained steered molecular dynamics model of a representative fibril during destructive tensile testing. The AGEs are inserted randomly between neighboring TC molecules in their helical regions. The model is based on our previous studies (Kamml et al., 2023b,a), where force-field parameters are obtained from (Depalle et al., 2015, 2016). The specificity of the collagen fibrils considered here is their mineral content, for which the modeling approach is described in detail below.

### 2.1. Geometry implementation of the collagen fibril

We build the geometry of the mineralized collagen fibril to represent the biological configuration of collagen type I including mineral in the fibrillar structure. First, we create the collagen fibril geometry *without* mineralization, where we use the same approach as described in (Kamml et al., 2023b,a). The TC molecules are arranged in the collagen-specific 5-staggering pattern with gap and overlap zones with a periodicity of $D = 67$ *nm*, with a gap size of $0.6 \cdot D$ and an overlap of $0.4 \cdot D$ (see Fig. 1a). The TC molecules are represented by a string of particles, where the bonded interactions between these particles represent the behavior of the TC molecules following a coarse-grained molecular dynamics approach (Buehler, 2006a,b; Depalle et al., 2015, 2016). The geometry of these TC molecules is obtained from Protein Data Bank entry 3HR2, the atomistic structure of TC extracted by X-ray crystallography (Orgel et al., 2006), and 218 particles are placed equidistantly along its longitudinal backbone. The fibril consists of 155 TC molecules per cross-section, resulting in a diameter of 200 nm. For smooth force transmission during tensile testing, the ends of the fibril are

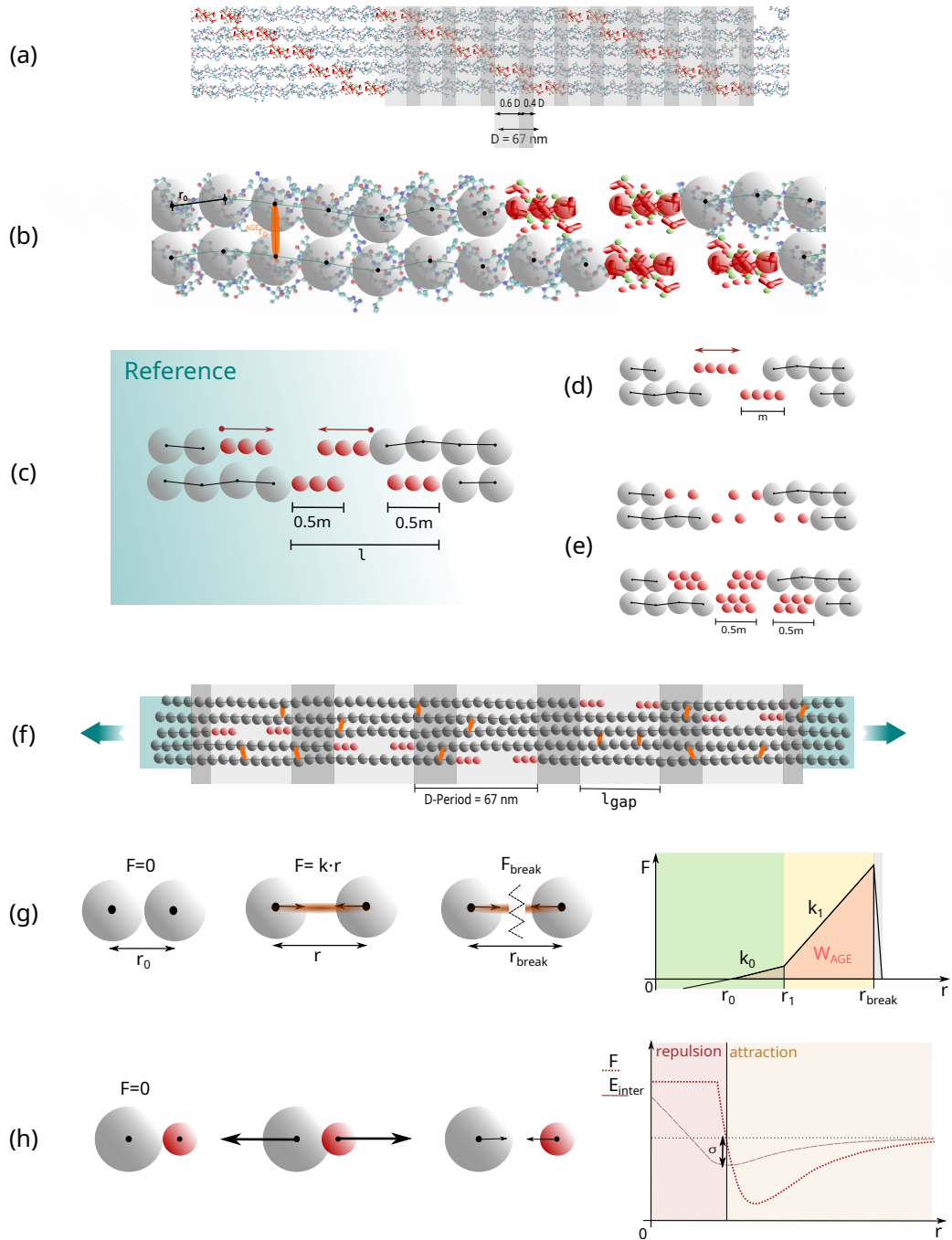

**Figure 1:** Schematic overview of model implementation and evaluation of deformation mechanisms. (a) Hodge-Petruska Model for collagen fibril: Displaying characteristic banding pattern with gap and overlap zones of TC molecules. (b) Coarse grained molecular dynamics model: The mechanical behavior of the TC molecules is represented by a string of particles mimicking the mechanical response that has been extracted from full scale simulations (Depalle et al., 2015). Mineral particles are shown in red, and collagen in grey. AGEs cross-links orange. (c) Schematic representation of insertion of mineral particles: mineral content $c_{mineral}$ is defined as $\frac{m}{l} \cdot 100\,\%$. Red arrows show the direction of mineralization. (d) Changes in nucleation of minerals: Mineralization starts from the center of the gap region. (e) Changes in the mineral pattern. Decreasing and increasing mineral particle density. (f) Schematic representation of our implementation of a representative collagen fibril geometry: 5 gap and overlap zones; AGEs cross-links were randomly inserted between TC molecules (red) with different densities; fibril is strengthened at the ends (blue area) to guarantee smooth force transmission. (g) Definition of bond interactions (collagen bonds in TC molecules and AGEs cross-links): trilinear bond behavior, where the force depends on the distance $r$ between two particles. The varied parameters in our simulations are $k_1$ and $r_{break}$; the loading energy capacity of a single bond, e.g., $W_{AGE}$ indicated by orange area. (h) Definition of non-bonded particle interactions due to van-der Waals forces: Soft-core Lennard-Jones Potential.

J. Kamml et al.  Page 3 of 13

extended with 40 particles per TC molecule and the bonds between these particles are strengthened. After creating the geometry of this collagen type I fibril, the mineralization is performed.

## 2.2. Mineralization of the collagen fibril

We insert mineral particles into the model to account for the mineralization of collagen fibrils in hard tissues like bone. The question of where mineral *i.e.* HAP-crystal, is exactly located in bone has not been answered to date and different opinions and theories exist at this point (see Discussion 4.1). Some claim that it is intrafibrillarly located between the TC molecules and in the gap zones (Arsenault, 1989; Siperko and Landis, 2001; Nudelman et al., 2010), while other experimental studies have shown that the mineral is also located extrafibrillar between the collagen fibrils (McNally et al., 2012; Tong et al., 2003; Landis et al., 1996; Su et al., 2003).

In our approach, the mineralization of the fibrils starts from the ends of the TC molecules at the sides of the gap zones and does not extend into the overlap region. This is similar to the approach of Nair et al. (2013), who performed full-atomistic simulations, with the mineral nearly exclusively located in the gap zones. This is in agreement with their experimental studies showing that the mineral phase is nucleated in the first section of the gap region after the transition from overlap to gap zones. We mineralize the gap zones by equidistantly placing HAP-particles starting from each overlap/gap transition point towards the inside of the gap (see Fig. 1c) at the equilibrium distance $r_0^{HAP}$ ($r_0^{HAP} = 2^{1/6} \cdot \sigma_{HAP}$). The distance at the collagen/mineral transition (between the last collagen and the first mineral particle) is the equilibrium distance $r_0^{col-HAP}$ ($r_0^{col-HAP} = 2^{1/6} \cdot \sigma_{col-HAP}$) where forces are 0, following the Lennard-Jones-Potential between collagen and HAP (see Fig. 1c, h and Tab. 1). This mineralization pattern is considered as the reference state of our mineralized collagen fibril. In addition, we created other models where mineral insertion was started from the center of the gap zones (see Fig. 1d) to evaluate the influence of the mineral nucleation position. Even more, for investigating how the morphology of minerals influences fibril mechanics, models with less mineral density were implemented. In these cases, either only every second particle position with respect to the reference state was occupied or two lines of mineral particles in an equidistant scaffold were added (see Fig. 1e). The mineral content is defined via the percentage of gap length that is occupied by mineral particles $c_{mineral} = m/l \cdot 100\%$ (see Fig. 1c-e).

## 2.3. Insertion of AGEs cross-links

The insertion process of AGEs cross-links is consistent with the one used in Kamml et al. (2023a,b). AGEs are inserted into the mineralized collagen fibril after the first equilibration of 20 ns. Aside from computational studies concentrating on individual AGEs (Collier et al., 2015; Gautieri et al., 2014), their exact location and where they act as cross-linking or non-cross-linking AGEs is unknown.

Hence, we insert them randomly between the central 95% helical regions of the collagen fibril. The AGEs content $N_{AGE}$ is measured per TC molecules, where AGEs are inserted per TC molecules to avoid accumulation effects. Further, since AGEs types and contents have not been quantified in bone, the content we apply in our models cannot be verified with experimental data, and, hence, we vary the content to investigate its effect.

## 2.4. Definition of the particle interactions

We use coarse-grained Molecular Dynamics for simulating the destructive tensile tests on the collagen fibrils. The total energy of the system is defined as the sum of all force field terms. For our mineralized collagen fibril, this translates to

$$\begin{aligned} E_{total} &= E_{bond} + E_{angle} + E_{non-bonded} \\ &= \sum_{bond} \Phi_{bond}(r) + \sum_{angle} \Phi_{angle}(\phi) + \\ &\quad + \sum_{non-bonded} \Phi_{non-bonded}(r) \,, \end{aligned} \quad (1)$$

where $E_{bond}$ is the bond energy due to stretching, $E_{angle}$ the dihedral bond interactions energy due to bending, and $E_{non-bonded}$ the pairwise interaction energy due to molecular interactions such as Van-der-Waals forces.

The interactions between particles are expressed through forces, and these forces between particles are calculated via particle distance

$$F = -\frac{\partial \Phi(r)}{\partial r} \quad (2)$$

or angle

$$F = -\frac{\partial \Phi(\phi)}{\partial \phi} \quad (3)$$

as the negative derivative of the potential energy.

The bond energy $E_{bond}$ includes collagen bonds between the particles of the TC molecules and AGEs cross-links. The bonds are modeled as trilinear springs with a regularization after bond breakage

$$F_{bond}(r) = \begin{cases} -k_T^{(0)}(r - r_0) & \text{if } r < r_1 \\ -k_T^{(1)}(r - r_0) & \text{if } r_1 \leq r < r_{break} \\ z \cdot k_T^{(1)}(r - r_0) & \text{if } r_{break} \leq r < r_{break} + a \\ 0 & \text{if } r \geq r_{break} + a \end{cases} \quad (4)$$

where $F_{bond}$ is the force acting between two particles that are connected with a bond, $k_T^{(0)}$ and $k_t^{(1)}$ are the respective spring constants of the bond deformation, $r_0$ is the equilibrium distance between the two bond particles and $a$ is defined as $a = z \cdot (r_{break} - r_1)$, with $z$ as the regularization factor to avoid any discontinuities and provide computational stability (see Fig. 1d).

We account for angle bending between a set of three particles with forces defined as

$$F_{angle}(\phi) = -k_B(\phi - \phi_i) \cdot \phi \,, \quad (5)$$

where $\phi_i$ are the varying equilibrium angles obtained from the initial TC molecule geometry, and $k_B$ is the bending stiffness of the molecule (Depalle et al., 2015; Buehler and Wong, 2006; Buehler, 2006b).

Van-der-Waals forces $E_{inter}$ define the potential between the different kinds of non-bonded particles: The collagen-to-collagen ($col$) interaction, the mineral-mineral ($HAP$) interaction and the collagen-mineral interaction ($col - HAP$) are modeled via a Lennard-Jones-Potential with a soft core following

$$F_{\text{non-bond}}(r) = \begin{cases} F_{LJ}(r) & \text{if } r \geq \lambda \sigma_{LJ} \\ F_{LJ}(\lambda \sigma_{LJ}) & \text{if } r < \lambda \sigma_{LJ} \end{cases} \quad (6)$$

where

$$F_{LJ}(r) = \frac{1}{r}\left[48\epsilon_{LJ}\left(\frac{\sigma_{LJ}}{r}\right)^{12} - 24\epsilon_{LJ}\left(\frac{\sigma_{LJ}}{r}\right)^6\right]. \quad (7)$$

with $\epsilon_{LJ}$ as the well depth between two particles, $\sigma_{LJ}$ is the distance at which the intermolecular potential between the two particles is zero and $\lambda$ is the parameter to adjust the critical force associated to the soft core (see Fig. 1e).

The parameters applied in our simulations for calculating particle interaction are displayed in Tab. 1. For cross-link modeling, we use the mechanical properties of glucosepane, extracted from full-atomistic simulations with a reactive force field, since glucosepane is the most abundant cross-link in tissue and has been shown to influence mechanics of non-mineralized collagen fibrils (Kamml et al., 2023a). Additionally, to invetigate the influence of AGEs cross-link mechanics, we increased the stiffness by a factor of 2.

### 2.5. Simulations

We perform destructive tensile tests on the models of the mineralized collagen fibrils in *LAMMPS* (Plimpton, 1995). In the first step, different contents of mineral are inserted into the gaps of the collagen fibrils to account for mineralization, followed by an equilibration for 20 ns in an NPT ensemble (300 K, 0 Pa) simulating an infinitely long fibril with periodic boundary conditions. After the insertion of cross-links, the tensile tests are performed, using steered molecular dynamics at a constant velocity of 0.0001 Å/fs (= 10 m/s) in an NVT ensemble at a temperature of 300 K. The time step is $\Delta t = 10$ fs in equilibration and tensile test simulation. The ends are moved apart, and the required force is measured to calculate the engineering stress within the collagen fibril.

## 3. Results

### 3.1. Effect of mineral on collagen fibril deformation behavior and strength

With our first set of simulations, we investigate the effect of changing mineral content in the gap zone, expressed in gap-filling percentages (see Fig. 1c – reference state). We observe that the mechanical behavior of the collagen fibril changes drastically in various aspects with increasing mineral content (see Fig. 2a). All fibrils show the same elastic behavior for strains up to $\varepsilon \approx 0.1$ independent of their mineral content. Fibrils with low contents, *i.e.* $c_{mineral} \leq 10\%$, present a softening behavior after reaching the peak stress $\sigma_{\text{peak}}$, eventually leading to zero strength. This behavior results from sliding between TC molecules, which is manifested in the observation that the collagen bonds within the TC molecules do not break (see Fig. 2b). The fibril with mineral content $c_{mineral} = 15\%$ presents a different behavior: after reaching the limit of linear deformation at $\varepsilon_0 \approx 0.15$ the modulus decreases. After reaching peaks stress $\sigma_{peak}$, it softens, and the behavior is characterized by a unique phenomenon within mineralized collagen fibrils – the so-called "sawtooth" effect (Depalle et al., 2016): The stress-strain curve is not smooth, but presents multiple relatively large stress-drops. These drops coincide with a slightly undulated behavior (see Fig. 2c) of the average gap length $l_{gap}$, as defined in Fig. 1f, which is the length of the gap zone between the particles at the ends of two TC molecules. At the same time, the number of broken collagen bonds remains relatively constant (see Fig. 2b), which suggests that this sawtooth behavior is the result of (dynamic) sliding events between the TC molecules.

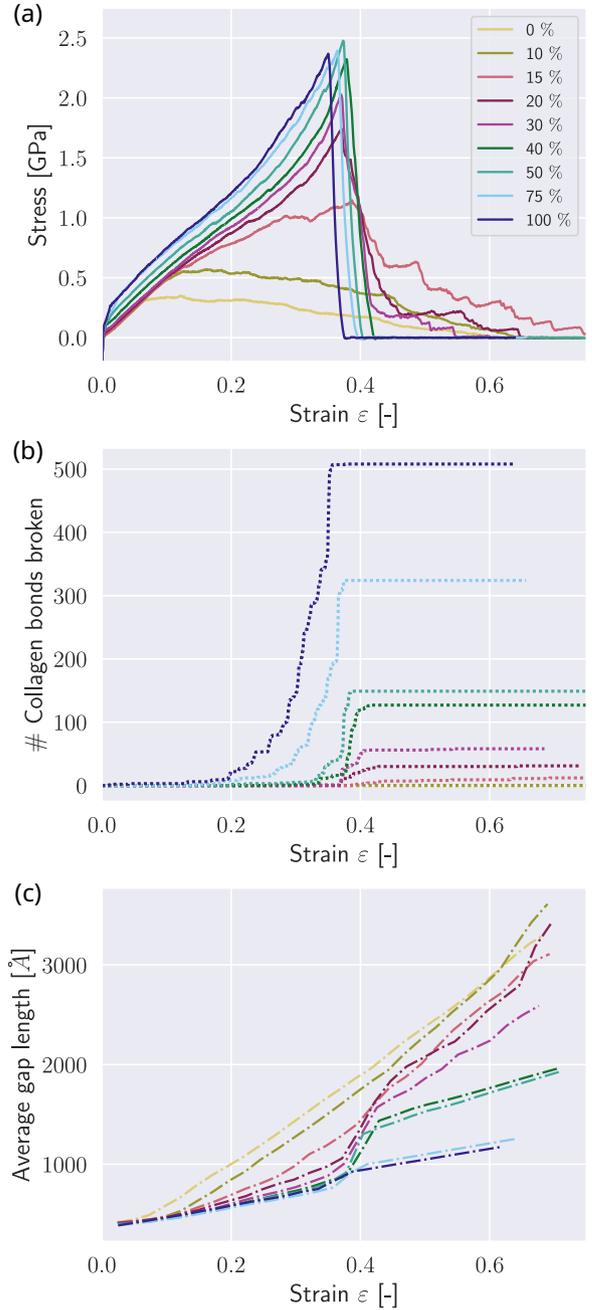

**Figure 2:** Mechanics of mineralized collagen fibril with varying contents of mineral. (a) Stress-strain curves of tensile tests until rupture of a representative collagen fibril with different mineral contents $c_{mineral}$ in [%]. (b) Number of broken bonds in TC molecules at fibril strain. (c) Average gap length at fibril strain (the average gap length is the average distance between the particles at the end of the TC molecules at both sides of the gap).

**Table 1**
Parameters used in coarse-grained molecular dynamics mesoscale model of mineralized collagen fibrils (Depalle et al., 2015, 2016; Kamml et al., 2023b)

| Components | Parameters | Value |
|---|---|---|
| Collagen molecules | Equilibrium particle distance ($r_0$, Å) | 14.00 |
| | Critical hyperelastic distance ($r_1$, Å) | 18.20 |
| | Bond breaking distance ($r_{break}$, Å) | 21.00 |
| | Tensile stiffness parameter ($k_0$, kcal mol$^{-1}$Å$^{-2}$) | 17.13 |
| | Tensile stiffness parameter ($k_1$, kcal mol$^{-1}$ Å$^{-2}$) | 97.66 |
| | Regularization factor ($z$, -) | 0.05 |
| | Equilibrium angle ($\phi_0$, degree) | 170.0 to 180.0 |
| | Bending stiffness parameter ($k_b$, kcal mol$^{-1}$rad$^{-2}$) | 14.98 |
| | Dispersive parameter ($\varepsilon_{col}$, kcal mol$^{-1}$) | 6.87 |
| | Dispersive parameter ($\sigma_{col}$, Å) | 14.72 |
| | Soft core parameter ($\lambda$, -) | 0.9 |
| | Mass of each mesoscale particle, atomic mass units | 1548 |
| Particles at ends of TC molecules | same parameters as collagen molecules except: | |
| | Bond breaking distance ($r_{break}$, Å) | 70.00 |
| Hydroxyapatite | Dispersive parameter ($\varepsilon_{HAP}$, kcal mol$^{-1}$) | 106.7 |
| | Dispersive parameter ($\sigma_{HAP}$, Å) | 10.28 |
| | Soft core parameter ($\lambda$, -) | 0.9 |
| | Cutoff radius ($c_{HAP}$, Å) | 13.85 |
| | Mass of each mesoscale particle, atomic mass units | 1324 |
| Interaction between collagen and hydroxyapatite | Dispersive parameter ($\varepsilon_{col-HAP}$, kcal mol$^{-1}$) | 137.1 |
| | Dispersive parameter ($\sigma_{col-HAP}$, Å) | 9.88 |
| | Soft core parameter ($\lambda$, -) | 0.9 |
| | Cutoff radius ($c_{col-HAP}$, Å) | 20.00 |
| AGEs Cross-links | Equilibrium particle distance ($r_0$, Å) | 18.52 |
| | Critical hyperelastic distance ($r_1$, Å) | 22.72 |
| | Bond breaking distance ($r_{break}$, Å) | 31.72 |
| | Tensile stiffness parameter ($k_0$, kcal, mol$^{-1}$Å$^{-2}$) | 0.1 |
| | Tensile stiffness parameter ($k_1$, kcal, mol$^{-1}$Å$^{-2}$) | 8.00 − 16.0 |

Fibrils with mineral contents of $20\% \leq c_{mineral} \leq 40\%$ show a slight decrease in modulus after reaching the limit of the elastic behavior at $\varepsilon_0$, followed by stiffening with increasing stress, starting at $\varepsilon \approx 0.3$. These fibrils finally reach their peak stress at $\sigma_{peak} \approx 2.5$ MPa. Fibrils with $c_{mineral} \leq 30\%$ still show a sawtooth behavior, while at higher $c_{mineral}$, the stress drops from $\sigma_{peak}$ directly to 0, which is a sign of brittleness. We also observe that higher contents of minerals lead to more broken bonds in the TC molecules (see Fig. 2b), where high drops in stress generally appear when a larger amount of bonds breaks. The bond breaking within the TC molecules indicates that the mineral blocks the sliding between the TC molecules. In general, mineral makes the fibril stronger *i.e.* increases the peak stress $\sigma_{peak}$ up to a mineral content of $c_{mineral} \approx 40\%$ when $\sigma_{peak}$ reaches its saturation level. Mineral contents of $c_{mineral} \leq 40\%$ do not lead to a further increase in $\sigma_{peak}$. We also note that the strain at peak stress $\varepsilon_{peak}$ increases up to $c_{mineral} \approx 30\%$ and then does not show any further significant changes. All of these results show that changes of $c_{mineral}$ have a stronger influence at lower mineral contents ($c_{mineral} \leq 30$) than at higher content, where the behavior does not change significantly when the mineral contents change.

In the following, the fibril with $c_{mineral} = 0\%$ that neither contains mineral nor cross-links is considered as the state of pure collagen molecules sliding. When observing the changes in the average length of the gap zone $l_{gap}$, this state of pure sliding shows a constant change of $l_{gap}$. The slope of the curves is generally decreasing with increased $c_{mineral}$, indicating that the TC molecules within the fibrils are stretched. When $c_{mineral} \geq 20\%$, the failure of the fibril and the larger amounts of broken collagen bonds in TC molecules coincide with the onset of a sudden change in slope at $\varepsilon_{peak} \approx 0.3$. The higher $c_{mineral}$, the less pronounced the changed modulus. This is an indicator that at a mineral content of $c_{mineral} = 40\%$, the collagen molecules are barely sliding, but the fibril breaks via a condensed break of collagen molecules. This is another indicator for increased brittleness of the collagen fibril with increasing mineral content.

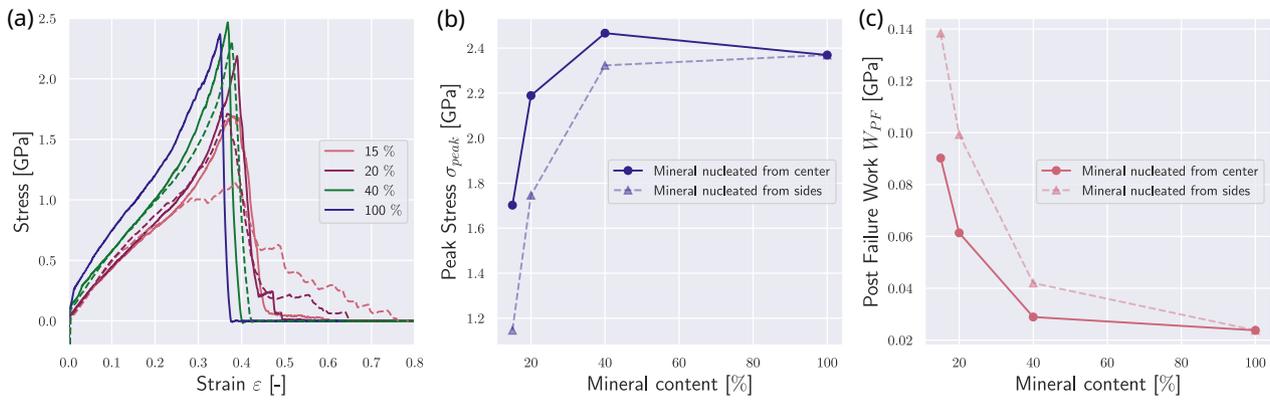

**Figure 3:** Effect of different mineral nucleation points on the mechanical response of the collagen fibril. Mineralization starting from the edges of the gap and growing towards the center are depicted with dashed lines, and correspond to the reference case shown in Fig. 2. Mineralization starting from the gap center and growing outwards are depicted with solid lines. (a) Stress-strain response of the collagen fibrils with different nucleation procedures. Mineral contents in [%]. (b) Comparison of peak stress $\sigma_\text{peak}$ in collagen fibrils with different mineral nucleation starting points. (c) Comparison of post-failure work $W_{PF}$ in collagen fibrils with different mineral nucleation starting points.

### 3.2. Effect of mineral nucleation position

We now investigate the effect of the position of the nucleation of the mineral. In the reference state presented in Sec. 3.1, nucleation of minerals starts at the respective ends of the gap zones. We compare this reference fibril to a fibril where the mineral nucleation starts in the center of the gap (see Fig. 1c). We observe that nucleation in the center intensifies the effect of the mineral on the collagen fibril mechanics. Specifically, the fibril demonstrates changes in mechanical behavior already at lower contents of the mineral when the mineral is nucleated from the center of the gap (see Fig. 3). With a mineral density of $c_{mineral} = 15\%$, the fibril with mineral extending from the center of the gap exhibits a behavior similar to the fibril with $c_{mineral} = 20\%$ in the reference fibril. The peak stress $\sigma_\text{peak}$ reaches its saturation level already at a mineral content of $c_{mineral} \approx 20\%$ (see Fig. 3a&b). Further, we consider the post-failure work $W_{PF}$, which is defined as

$$W_{PF} = \int_{\varepsilon_\text{peak}}^{\varepsilon_{PF}^0} \sigma(\varepsilon)\, d\varepsilon \qquad (8)$$

with $\varepsilon_\text{peak}$ corresponding to the strain at peak stress $\sigma_\text{peak}$ and $\varepsilon_{PF}^0$ being the strain when $\sigma \approx 0$ post fibril failure. We observe that $W_{PF}$ is generally smaller at the same mineral contents (see Fig. 3c). At the same $c_{mineral}$, the mineral nucleated from the center reaches higher stresses faster, but also loses the sawtooth behavior, leading to a more abrupt failure. With increasing mineral content, this difference is getting smaller, since the fibril has an increased $\sigma_\text{peak}$, but also more abrupt rupture, causing a decrease in $W_{PF}$. Since the nucleation position has mostly quantitative but no qualitative effects on the fibril behavior, we can infer that the mineral content is more important than its position but our results do not provide direct evidence on the location of mineral in the collagen fibril.

### 3.3. Effect of mineral pattern

To account for different forms and distributions of minerals, we additionally compare the reference fibrils' mechanical behavior with fibrils with other mineralization patterns. First, we decrease the mineral particle density in the gap zone, *i.e.*, only every second mineral particle position from the reference state is occupied (see Fig. 1e). In this configuration, the stress-strain curves show a decrease in the effect of minerals (see Fig. 4a). The stiffening is only initiated at higher mineral contents of about 40 %. We observe that the peak stress saturation level is not reached at mineral contents of $c_{mineral} \leq 40\%$, but only at $c_{mineral} \geq 75\%$ (see Fig. 4a&b). Due to reaching the level of abrupt failure only at very high mineral contents, the $W_{PF}$ is constantly high and only reaches its minimum at $c_{mineral} = 100\%$ (see Fig. 4c). This shows that when the mineral is not perfectly organized and defects, such as inclusions (here represented as empty mineral slots), exist, it still contributes quite efficiently to the strength of the collagen fibril but does not affect the fibril brittleness to the same degree.

Second, we increase the mineral particle density by adding minerals in an equidistant scaffold in two rows (see Fig. 1e). We observe the opposite effect, where the fibrils stiffen already at a mineral content of $c_{mineral} \approx 15\%$ and reach their peak stress saturation already with $c_{mineral} \approx 30\%$. At mineral contents of $c_{mineral} \approx 50\%$, we see a drop in $\sigma_\text{peak}$, and at $c_{mineral} \approx 100\%$, the stress-strain behavior does not significantly differ from the reference fibril.

J. Kamml et al.  Page 7 of 13

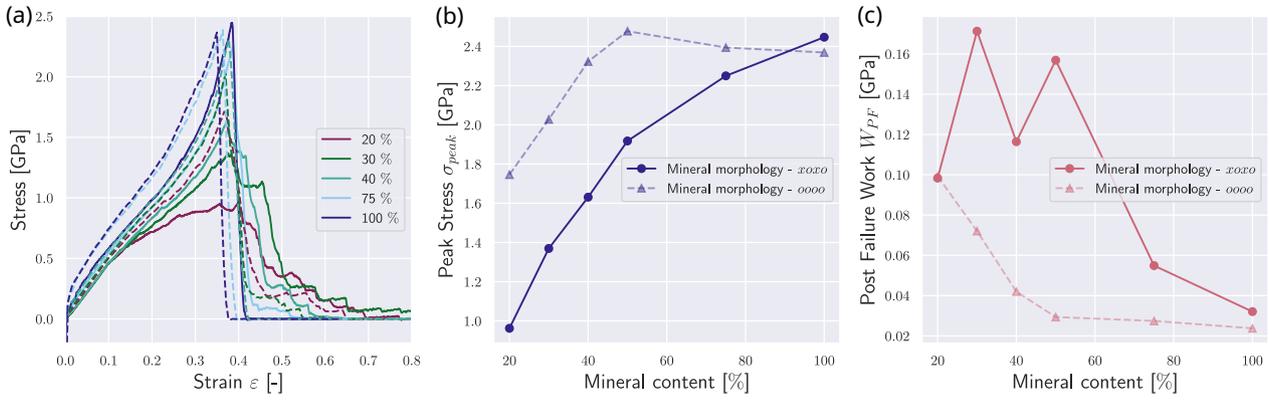

**Figure 4:** The influence of mineral morphology on the mechanical response of the collagen fibril. Results from simulations with mineralization patterns with one continuous line of minerals are depicted with dashed lines and correspond to the reference case shown in Fig. 2. Simulation results with mineralization where every second mineral position is empty are depicted with solid lines. (a) Stress-strain response of the collagen fibrils with different mineral morphologies. Mineral contents in [%]. (b) Comparison of peak stress $\sigma_{peak}$ in collagen fibrils with different mineral morphologies. (c) Comparison of post-failure work $W_{PF}$ in collagen fibrils with different mineral morphology.

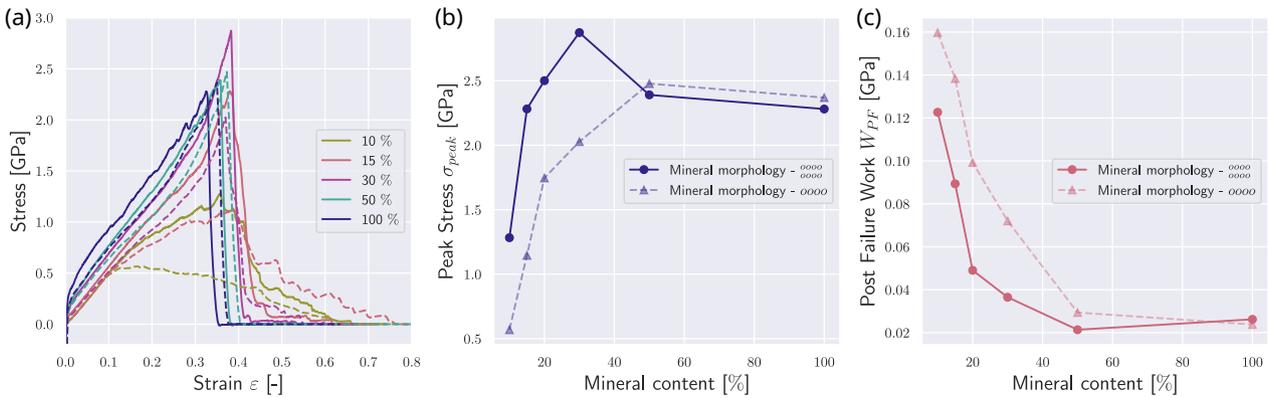

**Figure 5:** The influence of mineral morphology on the mechanical response of the collagen fibril: Mineralization pattern with one solid line of mineral (dashed lines) – reference case, see Fig 2 – and two solid lines of mineral (solid lines) (a) Stress-strain response of the collagen fibrils with different mineral morphologies. Mineral contents in [%]. (b) Comparison of peak stress $\sigma_{peak}$ in collagen fibrils with different mineral morphologies. (c) Comparison of post failure work $W_{PF}$ in collagen fibrils with different mineral morphology.

### 3.4. AGEs cross-linking in the mineralized collagen fibril

We investigate the influence of AGEs cross-linking on the collagen fibril mechanics at different mineral contents and varied not only AGEs density but additionally increased the stiffness of the AGEs by a factor of 2 to evaluate the influence of changing AGEs properties (see Fig. 6 and Tab. 1). Generally, when we observe AGEs and minerals separately, we observe that higher AGEs densities $N_{AGE}$ (without mineral) lead to higher peak stress $\sigma_{peak}$ and, depending on the AGEs properties, i.e. stiffness, leading to stiffening of the collagen fibril at the respective $N_{AGE}$ (at $k_1 = 8.0$ stiffening is initiated at $N_{AGE} = 40$, at $k_1 = 16.0$ at $N_{AGE} = 10$). The stiffening is initiated at $\varepsilon \approx 0.25$. At higher $N_{AGE}$ the failure of the fibril also is more abrupt, i.e. higher slope after reaching $\sigma_{peak}$. When observing the influence of $c_{mineral}$ (see Sec. 3.1), the mineral has similar effects with increasing peak stress $\sigma_{peak}$ and initiating a second stiffening. The $\sigma_{peak}$ reaches its saturation level at 2.5 GPa. The difference the stress-strain curve demonstrates is a reduction of the modulus after the linear regime. At this point, the subsequent stiffening is initiated but remains less pronounced for higher $c_{mineral}$. Further, the most striking difference is the sawtooth behavior: after reaching $\sigma_{peak}$, at a mineral content of $c_{mineral} \approx 15$. When $c_{mineral}$ is increased, the failure mechanism is more abrupt.

In all fibril configurations, the elastic modulus as observed from the initial fibril stiffness is independent of $c_{mineral}$ and $N_{AGE}$. Without minerals, AGEs cause the initiation of the second regime – the stiffening of the collagen fibril – which appears already at lower AGEs densities when the stiffness of the AGEs cross-links is increased (compare Fig. 6a&e).

When the mineral content $c_{mineral}$ of the fibril is at 15 %, the peak stress $\sigma_{peak}$ increases further for all $N_{AGE}$, when the stiffness of AGEs $k_1 = 8.0$. In case of $k_1 = 16.0$, $\sigma_{\text{peak}}$ only increases up to a $N_{AGE} \leq 10 AGEs/TC$. At this higher level of stiffness, the fibril reaches the saturation level of strength at $\sigma_{\text{peak}} = 4.0$ MPa. We observe that insertion of AGEs does not lead to the so-called sawtooth effect compared to insertion of mineral: while at lower mineral contents $c_{mineral} \leq 20\%$, we see also that $\sigma_{\text{peak}}$ increases with higher $c_{mineral}$, but the fibril shows this stick-slip behavior after having reached its peak stress. Insertion of AGEs at $N_{AGE} \geq 10$ AGEs/TC still leads to a more abrupt failure after reaching the same $\sigma_{\text{peak}}$. The higher the stiffness of the AGEs, the more pronounced the stiffening behavior of the fibril (compare Fig.6b&f).

At a mineral content of $c_{mineral} \geq 30\%$, the influence of AGEs density on the mechanical behavior of the fibril is reduced compared to $c_{mineral} = 15\%$ (see Fig. 6c, d, g, h). The fibril stiffens and fails abruptly, but the mineral at this stage is the dominating factor governing the deformation and failure behavior. Still, at higher $N_{AGE}$, the $\sigma_{\text{peak}}$ of the fibril is higher but reduces compared to the fibril at $c_{mineral} \leq 15$. When $c_{mineral} = 100\%$, we observe that AGEs do not affect the deformation behavior (see Fig. 6d&h).

Generally, we observe that AGE density influences collagen fibril mechanics at lower $c_{mineral}$, where the extent of manifestation is dependent on AGE stiffness, but when the mineral content increases to $c_{mineral} \geq 30$, the effect of AGEs reduces until no effect of AGEs can be observed (see Fig. 6d&h), independent of their stiffness.

When evaluating the breaking of bonds at different mineral contents $c_{mineral}$ and AGEs densities $N_{AGE}$, we observe that the higher the mineral content, the more mineral is dominating the mechanical response: While at low $c_{mineral}$, only at $N_{AGE} = 40$ AGES/TC collagen bonds in the TC molecules break and also just very few, the more mineral that is inserted, the more bonds in the collagen molecules break (see Fig. 7a-d). At $c_{mineral} = 100$, around 500 collagen bonds are broken. With 155 TC molecules per cross-section of the fibril, this means that, on average, more than three fracture sites occur within every TC molecule. When we observe the breaking of collagen bonds, we see the opposite effect (see Fig. 7e-h): with increasing $c_{mineral}$, the number of AGEs cross-links breaking decreases massively, from a maximum of 3800 cross-links in fibrils with $N_{AGE} = 40$ AGEs/TC at $c_{mineral} = 0\%$ to about 750 cross-links fractured at $c_{mineral} = 100\%$, which mean that at $c_{mineral} = 0\%$, more than 60 % of the cross-links break, compared to only about 12 % at $c_{mineral} = 100\%$. This indicates that at high mineral contents, the mineral in the gap zones acts like glue, attaching the molecules at the end of the TC molecules to the gap. This is in agreement with the measurements of the length of the gap zones in Fig. 8. We observe that at $c_{mineral} = 100\%$ the average gap length at a fibril strain of 0.5 is only one-third of the gap length at $c_{mineral} = 100\%$ (compare Fig. 8a&d). The average gap length constantly decreases with increasing $c_{mineral}$ (see Fig. 8a-d). Additionally, at increasing mineral, a sudden change of mineral length becomes more pronounced (see Fig. 8c) that finally disappears when the fibril reaches $c_{mineral} = 100\%$ where the average gap length is very low (see Fig. 8d).

## 4. Discussion
### 4.1. Limitations in fibril geometry and mineralization procedure

Our models of mineralized collagen fibrils allow us to observe the changes in mechanics and deformation behavior of the fibril as a function of changing mineral content and form and varying AGEs cross-link density. These models provide a groundwork for assessing the influence of cross-linking and mineralization. However, various other parameters and environmental conditions might contribute to collagen fibril mechanical behavior, and some of these factors have not yet been quantified or determined. For example, the form and distribution of minerals within the gap zones of the collagen fibrils have not been determined precisely, and the mineralization process concerning mineral nucleation, location, and distribution within and between collagen fibrils and fibers is a question of ongoing debate: McNally et al. (2012) used transmission electron microscopy to study the topological distribution and relation of collagen and minerals. They state that 70 % of the mineral occurs extrafibrillar being placed between and oriented parallel to the collagen fibril's longitudinal direction, in agreement with other studies (Tong et al., 2003; Landis et al., 1996; Su et al., 2003). The remaining 30 % are located in the gap zones. This is contradictory to other theories, stating that the mineral is also placed intrafibrillar between the TC molecules, extending to the overlap regions of the fibril (Arsenault, 1989; Siperko and Landis, 2001; Nudelman et al., 2010). Since the quantities and morphology of minerals have not been determined precisely, we used several different densities and patterns of mineral distributions in the gap zones. Still, by taking several factors and distributions into account, the fibril models are comprehensive enough to provide a reference for estimation of the influence of AGEs cross-links and mineral density in collagen fibrils in the future when parameters have been quantified.

Hydration has been shown to be an important factor regulating collagen fibril mechanics (Andriotis et al., 2023, 2014, 2019; Fielder and Nair, 2019), which we did not account for by modeling collagen fibrils without the influence of water molecules. The mechanical strength of cross-links is likely to be sensible to pH (Rennekamp et al., 2023), a fact that we do not account for in our simulations.

Although AGE cross-links occur in physical conditions only in the presence of enzymatic cross-links, we do not include them in the present study. We have shown in previous work that enzymatic cross-links do not significantly change the mechanical response of the collagen fibril (Kamml et al., 2023b). Still, it has been found that the function of enzymatic cross-links is more complex, especially of trivalent cross-links: one of their bonds might act as a sacrificial bond to maintain mechanical stability via energy dissipation and relaxation afterward, and bond breaking might activate tailored repair mechanisms before macroscopic failure (Rennekamp et al., 2023; Zapp et al., 2020). Apart from enzymatic cross-links, we did not account for different AGE types except by varying AGE cross-link stiffness since they have not been quantified in bone so far. Instead, we used the mechanical properties of glucosepane, the most abundant AGEs cross-link in tissue. Further, the distribution of AGEs was considered random since the exact locations of AGE cross-links are not known. When these locations are known, a future study should be performed on the influence of their locations.

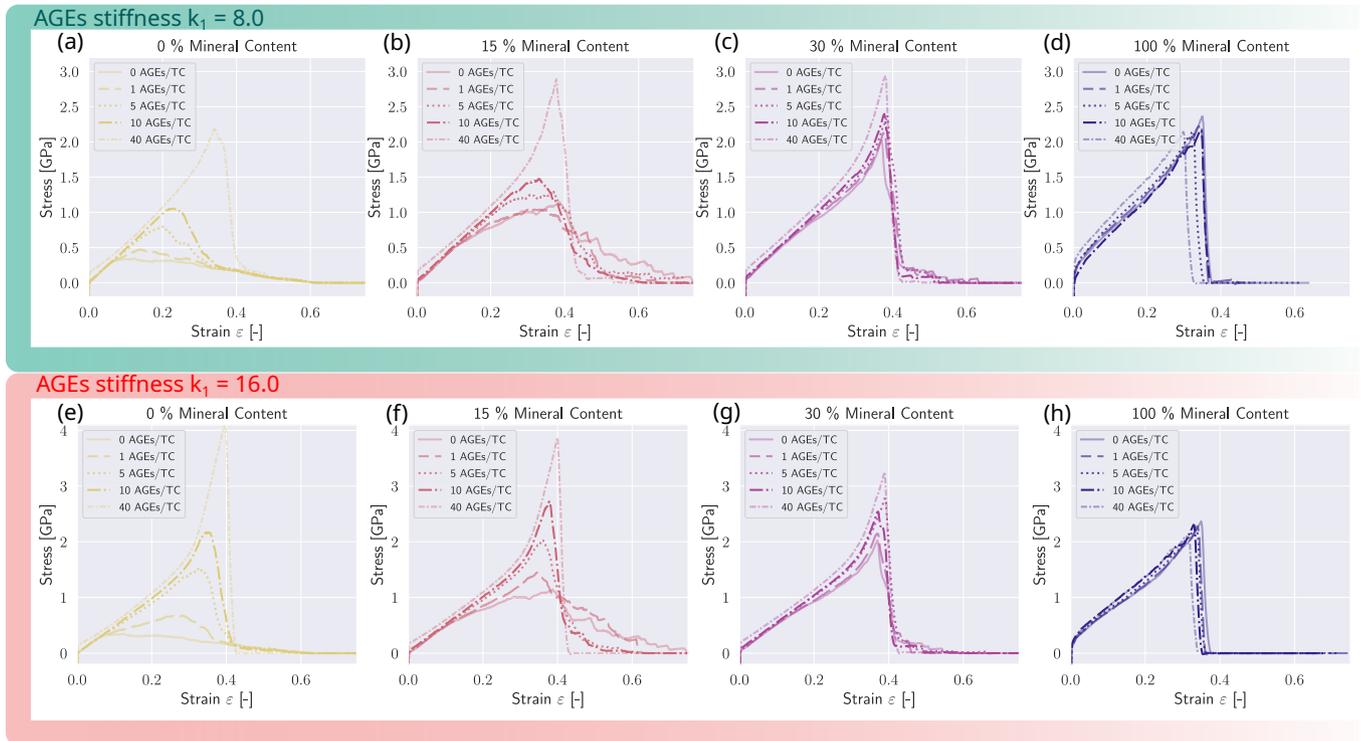

**Figure 6:** The influence of AGEs cross-link density and stiffness on the mechanical response of the mineralized collagen fibril. Stress-strain curves of mineralized collagen fibrils (and the non-mineralized fibril as reference (a&e)) at different mineral contents. (b&f) Mineral content of 15 %. (c&g) Mineral content of 30 %. (d&h) Mineral content of 100 %.

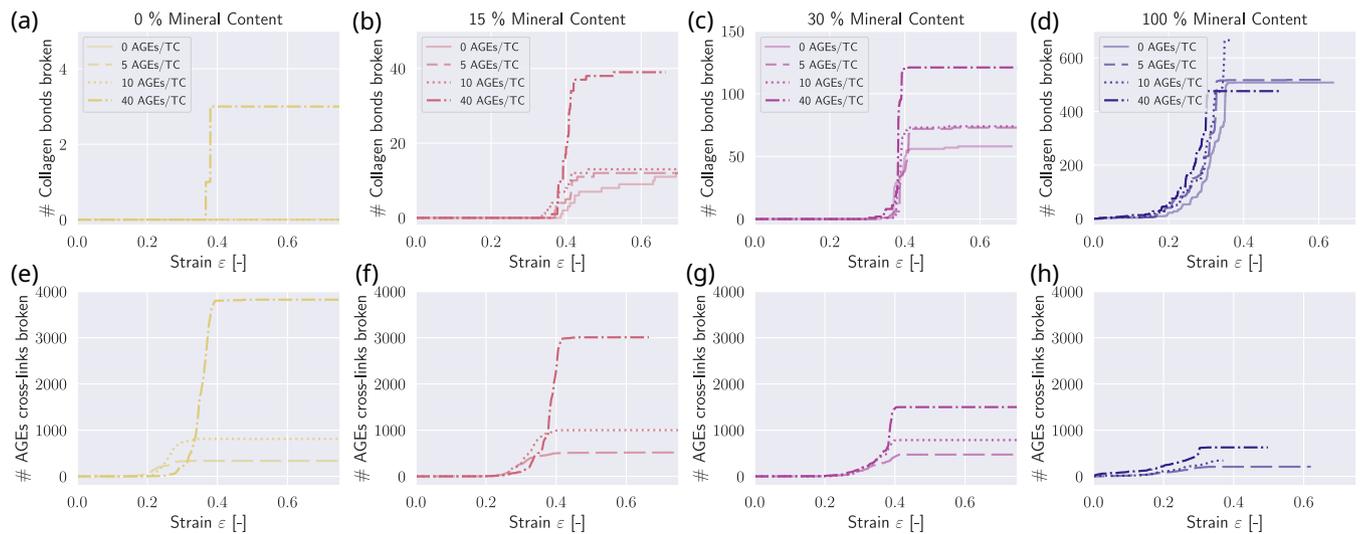

**Figure 7:** Development of bond breaking of collagen bonds within TC molecules and AGEs at various fibril strain levels with different AGEs densities $N_{AGE}$ at different mineral contents of $c_{mineral} = 0\%$ (a&e), $c_{mineral} = 15\%$ (b&f), $c_{mineral} = 30\%$ (c&g), and $c_{mineral} = 100\%$ (d&h). The tensile stiffness of the cross-links is $k_1 = 8.0$.

In addition to these factors, AGEs might not only influence the mechanical behavior of the collagen fibril via cross-linking but also via changing other important physiological functions like the hydration mechanism (Andriotis et al., 2019) or via imposing tissue resorption activities (Dong et al., 2011; Khosravi et al., 2014).

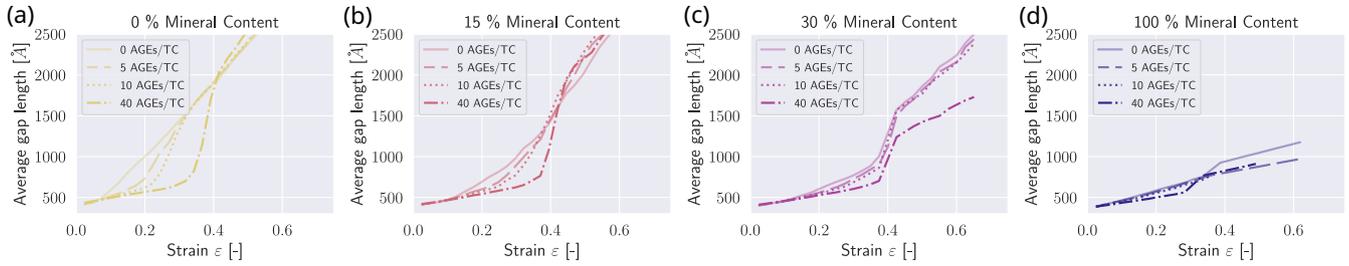

**Figure 8:** Measured length of gap region during tensile testing at different mineral contents $c_{mineral}$ including different AGEs densities. The tensile stiffness of the cross-links is $k_1 = 8.0$.

### 4.2. Implications of the results

Our results indicate that up to a certain content of mineral $c_{mineral}$, AGEs density $N_{AGE}$ influences the mechanical behavior ($c_{mineral} \leq 30\%$), but after the $c_{mineral}$ has exceeded this value, the mineral is dominating the deformation behavior. This leads to the conclusion that the deformation mechanism within the fibril changes: mineral-collagen interactions are strong enough to retain many particles at the gap/collagen transition (the last particles before the gap). This leads to a change in the failure mechanism, where a higher amount of collagen bonds breaks due to the retaining of the gap particles.

The increase of the peak stress $\sigma_{peak}$ up to a certain value and the fact that at higher stiffness $k_1$ at $N_{AGE} = 40$ it does not increase at $c_{mineral} = 15\%$ compared to $c_{mineral} = 0\%$ indicates that the transfer of forces to TC molecules via reduction of sliding reaches a saturation level where tropocollagen bonds break. The mineralized collagen fibril is probably a highly tuned biological system. Diabetes might not only influence collagen fibril mechanics via AGEs cross-linking but disturb the metabolic process of bone resorption (Montagnani et al., 2011): Indeed, studies show that the mineral content of bone in T2DM is increased (Hadzibegovic et al., 2008; Oei et al., 2013; Ma et al., 2012), raising the question of whether it is the increased AGEs cross-linking or the increased mineral content that is disturbing the tuned mechanical properties (or a combination of both). We observe that both increased AGE density and mineral content cause a more abrupt failure of the fibril. The stiffening of the fibrils is caused by changes in deformation mechanisms. The mineral content as well as AGEs block the sliding of the TC molecules. Finally, we note that further investigation on the presence of different AGEs types, *i.e.* AGEs with different mechanical properties, are needed for a comprehensive understanding of tissue mechanics at the nano-scale and for a more profound study on AGEs cross-linking in mineralized collagen fibrils.

### 5. Conclusion

With our model of the mineralized collagen fibril, we investigated the influence of both, mineral content and AGEs cross-link density on the mechanics of the fibril. We demonstrated that both of these components generally increase the strength of the fibril. Furthermore, increasing their quantity leads to the onset of another deformation regime characterized by stiffening of the fibril and resulting in a more abrupt failure. We showed that at higher mineral content, the mineral dominates the deformation and failure mechanism of the fibril. In contrast, at lower contents of minerals, AGEs dominate the mechanisms of deformation. These results indicate that distortions in either the deposition of mechanical components (*e.g.*, increased or reduced mineral content, or increased AGEs cross-linking) or of their mechanical properties alter the mechanical behavior of tissue on a larger scale by changing the collagen fibril mechanics.

### 6. Funding

Research reported in this publication was supported by NIAMS of the National Institutes of Health under award number 1R21AR077881.

### CRediT authorship contribution statement

**Julia Kamml:** Formal analysis, Investigation, Visualization, Writing - Original Draft . **Claire Acevedo:** Conceptualization, Funding acquisition, Writing - Review & Editing . **David S. Kammer:** Conceptualization, Methodology, Funding acquisition, Writing - Review & Editing .